\documentclass{article}%
\usepackage{amsfonts}
\usepackage{amsmath}
\usepackage{amssymb}
\usepackage{graphicx}%
\newtheorem{theorem}{Theorem}

\newtheorem{definition}[theorem]{Definition}

\newtheorem{lemma}[theorem]{Lemma}

\newtheorem{remark}[theorem]{Remark}

\newenvironment{proof}[1][Proof]{\textbf{#1.} }{\ \rule{0.5em}{0.5em}}
\begin{document}

\title{Dirac reduction revisited}
\author{Krzysztof MARCINIAK~$^{\dag}$ and Maciej B\L ASZAK~\thanks{Partially supported
by KBN grant No. 5P03B 004 20} $^{\ddag}$}
\date{$^{\dag}$ Department of Science and Technology \\
Campus Norrk\"{o}ping, Link\"{o}ping University\\
601-74 Norrk\"{o}ping, Sweden \\
~~E-mail: krzma@itn.liu.se\\
[10pt] $^{\ddag}$ Institute of Physics, A. Mickiewicz University\\
Umultowska 85, 61-614 Pozna\'{n}, Poland \\
~~E-mail: blaszakm@amu.edu.pl\\
[10pt]March 7, 2003}
\maketitle

\begin{abstract}
The procedure of Dirac reduction of Poisson operators on submanifolds is
discussed within a particularly useful special realization of the general
Marsden-Ratiu reduction procedure. The Dirac classification of constraints on
'first-class' constraints and 'second-class' constraints is reexamined.

\end{abstract}


\section{Introduction}

Dirac bracket as well as Dirac's classification of constraints is nowadays a
well recognized and very useful tool in the construction of Poisson dynamics
on admissible submanifolds from a given Poisson dynamics on a given manifold.
In this paper we consider the Dirac reduction procedure in a more general
setting than is usually met in literature. In Section 2 we implement the Dirac
reduction procedure into a particularly useful special realization of the
general Marsden-Ratiu reduction scheme, based on the concept of transversal
distributions. In Section 3 we reconsider the Dirac concept of first class
constraints as it seems to be too restrictive.

Firstly we recall few basic notions from Poisson geometry. Given a manifold
$\mathcal{M}$, a \emph{Poisson operator} $\pi$ on $\mathcal{M}$ is a mapping
$\pi:T^{\ast}\mathcal{M}\rightarrow T\mathcal{M}$ that is fibre-preserving
(i.e. $\pi|_{T_{x}^{\ast}\mathcal{M}}:T_{x}^{\ast}\mathcal{M}\rightarrow
T_{x}\mathcal{M}$ for any $x\in\mathcal{M}$) and such that the induced bracket
on the space $C^{\infty}(\mathcal{M})$ of all smooth real-valued functions on
$\mathcal{M}$
\begin{equation}
\left\{  .,.\right\}  _{\pi}:C^{\infty}(\mathcal{M})\times C^{\infty
}(\mathcal{M})\rightarrow C^{\infty}(\mathcal{M})\text{ \ , \ }\left\{
F,G\right\}  _{\pi}\overset{\mathrm{def}}{=}\left\langle dF,\pi
\,dG\right\rangle , \label{bracket}%
\end{equation}
where $\left\langle .,.\right\rangle $ is the dual map between $T\mathcal{M} $
and $T^{\ast}\mathcal{M}$, is skew-symmetric and satisfies Jacobi identity
(the bracket (\ref{bracket}) always satisfies the Leibniz rule $\left\{
F,GH\right\}  _{\pi}=G\left\{  F,H\right\}  _{\pi}+H\left\{  F,G\right\}
_{\pi}$). The symbol $d$ denotes the operator of exterior differentiation. The
operator $\pi$ can always be interpreted as a bivector, $\pi\in\Lambda
^{2}(\mathcal{M})$ and in a given coordinate system $(x^{1},\ldots,x^{m})$ on
$\mathcal{M}$ we have
\[
\pi=\sum\limits_{i<j}^{m}\pi^{ij}\frac{\partial}{\partial x_{i}}\wedge
\frac{\partial}{\partial x_{j}}.
\]
A function $C:\mathcal{M}\rightarrow\mathbb{R}$ is called a \emph{Casimir
function} of the Poisson operator $\pi$ if for an arbitrary function
$F:\mathcal{M}\rightarrow\mathbb{R}$ we have $\left\{  F,C\right\}  _{\pi}=0$
or, equivalently, if $\pi dC=0$.

\section{Marsden-Ratiu reduction for transversal distributions}

The Marsden-Ratiu reduction theorem \cite{MarsdenRatiu} describes the
procedure of reducing a Poisson operator $\pi$ on arbitrary submanifold
$\mathcal{S}$ of our manifold $\mathcal{M}$. This general procedure exists
only if some conditions are satisfied. These conditions involve a distribution
$E$ (in the original notation of Marsden and Ratiu) that is a subbundle of
$T\mathcal{M}$. By a simple assumption, namely that this distribution is
transversal, one can, however, satisfy all these conditions automatically.
Below we reformulate the Marsden-Ratiu theorem in this more limited but useful setting.

Consider an $m$-dimensional manifold $\mathcal{M}$ equipped with a Poisson
operator $\pi$ and an $s$-dimensional submanifold $\mathcal{S}$ of
$\mathcal{M}$. Fix a distribution $\mathcal{Z}$ of constant dimension $k=m-s$,
that is a smooth collection of $m$-dimensional subspaces $\mathcal{Z}%
_{x}\subset$ $T_{x}\mathcal{M}$ at every point $x$ in $\mathcal{M}$, which is
transversal to $\mathcal{S}$ in the sense that no vector field $Z\in
\mathcal{Z}$ is at any point tangent to the submanifold $\mathcal{S}$. Hence
we have
\[
T_{x}\mathcal{M}=T_{x}\mathcal{S}\oplus\mathcal{Z}_{x}%
\]
for every $x\in\mathcal{S}$ and, similarly,%
\[
\text{\ \ }T_{x}^{\ast}\mathcal{M}=T_{x}^{\ast}\mathcal{S}\oplus
\mathcal{Z}_{x}^{\ast},
\]
where $T_{x}^{\ast}\mathcal{S}$ is the annihilator of $\mathcal{Z}_{x}$ and
$\mathcal{Z}_{x}^{\ast}$ is the annihilator of $T_{x}\mathcal{S}$. That means
that if $\alpha$ is a one form in $T_{x}^{\ast}\mathcal{S}$ then $\alpha(Z)=0$
for all vectors $Z\in\mathcal{Z}_{x}$ and if $\beta$ is a one-form in
$\mathcal{Z}_{x}^{\ast}$ then $\beta$ vanishes on all vectors in
$T\mathcal{S}_{x}$.

\begin{definition}
A function $F:\mathcal{M}\rightarrow\mathbb{R}$ is invariant with respect to
$\mathcal{Z}$ \ if $L_{Z}F=Z(F)=0$ for any $Z\in\mathcal{Z}$. Similarly, a
function $F:\mathcal{M}\rightarrow\mathbb{R}$ is invariant with respect to
$\mathcal{Z}$ \ on $\mathcal{S}$ $\ $($\mathcal{Z}|_{\mathcal{S}}$-invariant
in short) if $L_{Z}F|_{\mathcal{S}}=Z(F)|_{\mathcal{S}}=0$ for any
$Z\in\mathcal{Z}$
\end{definition}

Here and in what follows the symbol $L_{Z}$ means the Lie derivative along the
vector field $Z$.

\begin{definition}
An operator $\pi$ is called invariant with respect to the distribution
$\mathcal{Z}$ \ if the functions that are $\mathcal{Z}$- invariant form a
Poisson subalgebra, that is, if $F$, $G:\mathcal{M}\rightarrow\mathbb{R}$ are
two $\mathcal{Z}$- invariant functions then $\left\{  F,G\right\}  _{\pi}$ is
again a $\mathcal{Z}$- invariant function. Similarly, an operator $\pi$ is
called invariant with respect to the distribution $\mathcal{Z}$ \ on
$\mathcal{S}$ if the functions that are $\mathcal{Z}|_{\mathcal{S}}$ -
invariant form a Poisson subalgebra, that is, if $F$, $G:\mathcal{M}%
\rightarrow\mathbb{R}$ are two $\mathcal{Z}|_{\mathcal{S}}$ - invariant
functions then $\left\{  F,G\right\}  _{\pi}$ is again a $\mathcal{Z}%
|_{\mathcal{S}}$ - invariant function.
\end{definition}

We denote these Poisson subalgebras by $\mathcal{A}$ and $\mathcal{A}%
_{\mathcal{S}}$ respectively. Let us observe, that if $\pi$ is $\mathcal{Z}%
|_{\mathcal{S}_{\mathcal{\nu}}}$-invariant for any manifold $\mathcal{S}%
_{\mathcal{\nu}}$ in a foliation of $\mathcal{M}$ then it is also
$\mathcal{Z}$- invariant.

\begin{theorem}
(Marsden and Ratiu \cite{MarsdenRatiu}): Let $\mathcal{S}$ be a submanifold of
$\mathcal{M}$ equipped with a Poisson operator $\pi$ and let $\mathcal{Z}$ be
a distribution in $\mathcal{M}$ that is transversal to $\mathcal{S}$. If the
operator $\pi$ is invariant with respect to the distribution $\mathcal{Z}$ on
$\mathcal{S}$, then the Poisson operator $\pi$ is reducible on $S$ in the
sense that on $S$ there exists a (uniquely defined) Poisson operator $\pi_{R}$
such that for any $f,g:S\rightarrow\mathbb{R}$ we have%
\begin{equation}
\left\{  f,g\right\}  _{\pi_{R}}=\left\{  F,G\right\}  _{\pi}|_{S}
\label{redukcja}%
\end{equation}
for any $\mathcal{Z}|_{\mathcal{S}}$ - invariant prolongations $F$ and $G$ of
$f$ and $g$ respectively.
\end{theorem}

The above construction, however, is difficult to perform in practice since it
is often hard to find explicit expressions for the prolongations $F$ and $G$.
We now show how this difficulty can be omitted.

Firstly, suppose that our submanifold $\mathcal{S}$ is given by $k$
functionally independent equations $\varphi_{i}(x)=0$, $i=1,\ldots,k$
(constraints) and that our transversal distribution $\mathcal{Z}$ is spanned
by $k$ vector fields $Z_{i}$ chosen such that the following orthogonality
relation holds%
\begin{equation}
\left\langle d\varphi_{i},Z_{j}\right\rangle =Z_{j}(\varphi_{i})=\delta_{ij},
\label{ortogonalnosc}%
\end{equation}
(this is no restriction since for any distribution $\mathcal{Z}$ transversal
to $\mathcal{S}$ we can choose its basis so that (\ref{ortogonalnosc}) is
satisfied). We observe that in this case we have $\left[  Z_{i},Z_{j}\right]
\varphi_{k}=0$ for all $k$, where $[X,Y]=L_{X}Y=X(Y)-Y(X)$ is the Lie bracket
(commutator) of the vector fields $X,Y$, so that $\left[  Z_{i},Z_{j}\right]
$ is always tangent to $\mathcal{S}$. Then, in case that the distribution
$\mathcal{Z}$ is involutive (integrable), this means that $\left[  Z_{i}%
,Z_{j}\right]  =0$ for all $i,j$. Moreover, we define the vector fields
$X_{i}$ as%
\begin{equation}
X_{i}=\pi d\varphi_{i}\text{, \ }i=1,\ldots,k. \label{xi}%
\end{equation}

There exists an important class of $\mathcal{Z}$-invariant Poisson operators

\begin{lemma}
\label{Vaismanwar}\cite{Vaisman} If
\begin{equation}
L_{Z_{i}}\pi=%
{\displaystyle\sum\limits_{j=1}^{k}}
W_{j}^{(i)}\wedge Z_{j}\text{ \ \ }i=1,\ldots,k \label{Vais}%
\end{equation}
for some vector fields $W_{j}^{(i)}$, then the Poisson operator $\pi$ is
invariant with respect to $\mathcal{Z}$
\end{lemma}

We sketch the proof here for the clarity of the text.

\begin{proof}
Assume, that $L_{Z_{i}}F=L_{Z_{i}}G=0$ for all $i$. We have to show that
$L_{Z_{i}}\left\{  F,G\right\}  _{\pi}=0$ for all $i$, but, due to
(\ref{Vais})%
\[
L_{Z_{i}}\left\{  F,G\right\}  _{\pi}=L_{Z_{i}}\left\langle dF,\pi
dG\right\rangle =%
{\displaystyle\sum\limits_{j=1}^{k}}
\left\langle dF,(W_{j}^{(i)}\wedge Z_{j})dG\right\rangle
\]
since $L_{Z_{i}}(dF)=d(L_{Z_{i}}F)=0$ (and similarly for $G$). On the other
hand%
\[
\left\langle dF,(W_{j}^{(i)}\wedge Z_{j})dG\right\rangle =Z_{j}(G)W_{j}%
^{(i)}(F)-Z_{j}(F)W_{j}^{(i)}(G)=0
\]
since $Z_{j}(F)=L_{Z_{j}}F=0$ (and similarly for $G$).
\end{proof}

The condition (\ref{Vais}) is sufficient but not necessary. For example, if
\[
L_{Z_{i}}\pi=%
{\displaystyle\sum\limits_{j=1}^{k}}
W_{j}\wedge\left[  Z_{i},Z_{j}\right]  \text{ \ \ }i=1,\ldots,k
\]
for some vector fields $W_{i}$, then the operator $\pi$ is also $\mathcal{Z}%
$-invariant (one shows it by computations similar to those in the above
proof). In the case when $\pi$ satisfies (\ref{Vais}) we apply the Lie
derivative $L_{Z_{j}}$ to both sides of the equation (\ref{xi}). Due to
(\ref{Vais}) we obtain%
\begin{align}
\left[  Z_{j},X_{i}\right]   &  =L_{Z_{j}}X_{i}=(L_{Z_{j}}\pi)d\varphi
_{i}=\left(  \sum_{l}W_{l}^{(j)}\wedge Z_{l}\right)  d\varphi_{i}=\nonumber\\
& \label{zwiazek}\\
&  =\sum_{l}\left(  Z_{l}(\varphi_{i})W_{l}^{(j)}-W_{l}^{(j)}(\varphi
_{i})Z_{l}\right)  =W_{i}^{(j)}-\sum_{l}W_{l}^{(j)}(\varphi_{i})Z_{l}\nonumber
\end{align}

We observe that, if $F$ and $G$ are two $\mathcal{Z}|_{\mathcal{S}}$-invariant
functions and $V_{j}$ are arbitrary vector fields, then $\left.  \left\langle
dF,%
{\textstyle\sum_{j}}
V_{j}\wedge Z_{j}\text{ }dG\right\rangle \right\vert _{\mathcal{S}}=0$ since
$\left\langle dF,V_{j}\wedge Z_{j}\text{ }dG\right\rangle =$ $Z_{j}%
(G)V_{j}(F)-Z_{j}(F)V_{j}(G)=0$ on $\mathcal{S}$. Thus the Poisson operator
$\pi$ and its \emph{deformation} of the form
\begin{equation}
\pi_{D}=\pi-%
{\textstyle\sum_{j}}
V_{j}\wedge Z_{j} \label{deformacja}%
\end{equation}
in spite of the fact that they act differently on $\mathcal{A}_{\mathcal{S}}$
both generate the same bracket on $\mathcal{S}$ so that both can be used to
define our restricted operator $\pi_{R}$ on $\mathcal{S}$ through
(\ref{redukcja}). Of course, the deformed operator $\pi_{D}$ does not have to
be Poisson, but nevertheless its restriction to $\mathcal{S}$ through
(\ref{redukcja}) must be Poisson since it naturally coincides with similar
restriction of $\pi$ to $\mathcal{S}$. If we now consider a whole foliation of
$\mathcal{M}$ defined by the functions $\varphi_{i}$ with leaves
$\mathcal{S}_{\nu}$ (so that $\mathcal{S}_{0}=\mathcal{S}$) then it turns out
that we can choose our (undetermined so far) vector fields $V_{j}$ in
(\ref{deformacja}) so that%
\begin{equation}
\pi_{D}(\alpha_{x})\in T_{x}\mathcal{S}_{\nu}\text{ \ \ for any }\alpha_{x}\in
T_{x}^{\ast}\mathcal{M}\text{ and any }x\in\mathcal{M}, \label{stycznosc}%
\end{equation}
which has a far reaching consequence.

\begin{lemma}
The deformation $\pi_{D}$ given by (\ref{deformacja}) that also satisfies
(\ref{stycznosc}) is Poisson.
\end{lemma}

\begin{proof}
The condition that $\pi_{D}(\alpha_{x})$ is tangent to $\mathcal{S}_{\nu}$ for
any $\alpha_{x}\in T_{x}^{\ast}\mathcal{M}$ is equivalent to the requirement
that $\left\langle d\varphi_{i},\pi_{D}(\alpha_{x})\right\rangle =0$ for all
$i$. Due to the antisymmetry of $\pi_{D}$ this requirement can be rewritten as
$\left\langle \alpha_{x},\pi_{D}(\varphi_{i})\right\rangle =0$ for all $i$.
Since $\alpha_{x}$ is arbitrary, the condition attains the form $\pi
_{D}(d\varphi_{i})=0$ for $i=1,\ldots,k$. We now complete the set of functions
$\varphi_{i}$ with some functions $x_{j}$ to a coordinate system $(x,\varphi)$
on $\mathcal{M}$. Then the matrix of the operator $\pi_{D}$ has the last $k$
rows and last $k$ columns equal to zero while the $m-k$ dimensional upper left
block coincides on every leaf $\mathcal{S}_{\nu}$ with the correponding
$\pi_{R}$ which is Poisson by the Marsden-Ratiu construction.
\end{proof}

\begin{lemma}
The condition (\ref{stycznosc}) can be written as%
\begin{equation}
V_{i}-\sum_{j=1}^{k}V_{j}(\varphi_{i})Z_{j}=X_{i} . \label{warunek}%
\end{equation}

\end{lemma}

\begin{proof}
We know that the condition (\ref{stycznosc}) can be written as $\pi
_{D}(d\varphi_{i})=0$ for $i=1,\ldots,k$. An easy calculation yields now that%
\begin{align*}
0  &  =\pi_{D}(d\varphi_{i})=\pi(d\varphi_{i})-\sum_{j=1}^{k}\left(
Z_{j}(\varphi_{i})V_{j}-V_{j}(\varphi_{i})Z_{j}\right)  =\\
&  =X_{i}-V_{i}+\sum_{j=1}^{k}V_{j}(\varphi_{i})Z_{j}%
\end{align*}
due to the normalization condition (\ref{ortogonalnosc}).
\end{proof}

We now restrict ourselves to only two limit cases, when all $X_{i}$ are
tangent to $\mathcal{S}$ and when $X_{i}$ span $\mathcal{Z}$.

\subsection{The case when $X_{i}$ are tangent to $\mathcal{S}$}

We firstly assume that all the vectors $X_{i}$ are tangent to $\mathcal{S}$.
We have then naturally $X_{i}(\varphi_{j})=0$. This in turn means that
$\left\{  \varphi_{i},\varphi_{j}\right\}  _{\pi}=\left\langle d\varphi
_{i},\pi d\varphi_{j}\right\rangle =\left\langle d\varphi_{i},X_{j}%
\right\rangle =0$ so that all the vector fields $X_{i}$ commute. In this case
the simplest solution of (\ref{warunek}) has the form $V_{i}=X_{i}$ and the
corresponding deformation (\ref{deformacja}) attains the form%
\begin{equation}
\pi_{D}=\pi-\sum_{i=1}^{k}X_{i}\wedge Z_{i}. \label{defstyczna}%
\end{equation}
This deformation has been recently widely used for projecting Poisson pencils
on symplectic leaves of one of their operators \cite{pedroni}-\cite{m4}.

\begin{lemma}
\cite{pedroni} The vector fields $W_{j}^{(k)}$ in (\ref{Vais}) can, in the
case that all $X_{i}$ are tangent to $\mathcal{S}$, be chosen as tangent to
$\mathcal{S}$.
\end{lemma}

\begin{proof}
Consider the projections $\widetilde{W}_{j}^{(i)}$of the vector fields
$W_{j}^{(i)}$ onto $\mathcal{S}$:%
\[
\widetilde{W}_{j}^{(i)}=W_{j}^{(i)}-\sum_{r=1}^{k}W_{j}^{(i)}(\varphi
_{r})Z_{r}.
\]
If $W_{j}^{(i)}$ are in $\mathcal{Z}$, then $\widetilde{W}_{j}^{(i)}=0$. \ The
vector field $\widetilde{W}_{j}^{(i)}$ is indeed tangent to $\mathcal{S}$
since
\[
\widetilde{W}_{j}^{(i)}(\varphi_{l})=W_{j}^{(i)}(\varphi_{l})-\sum_{r=1}%
^{k}W_{j}^{(i)}(\varphi_{r})\delta_{lr}=0.
\]
Now
\[
\sum_{j=1}^{k}\widetilde{W}_{j}^{(i)}\wedge Z_{j}=\sum_{j=1}^{k}W_{j}%
^{(i)}\wedge Z_{j}-\sum_{j,r=1}^{k}W_{j}^{(i)}(\varphi_{r})Z_{r}\wedge Z_{j}%
\]
the last term being equal to zero since $L_{Z_{k}}\left\{  \varphi_{i}%
,\varphi_{j}\right\}  _{\pi}=0$ implies $W_{j}^{(i)}(\varphi_{r})=W_{r}%
^{(i)}(\varphi_{j})$. Thus $\sum_{j=1}^{k}W_{j}^{(i)}\wedge Z_{j}=\sum
_{j=1}^{k}\widetilde{W}_{j}^{(i)}\wedge Z_{j}$.
\end{proof}

Due to this gauge freedom, if we choose $W_{j}^{(i)}$ as tangent to
$\mathcal{S}$ (which means that $W_{j}^{(i)}(\varphi_{r})=0$) then the formula
(\ref{zwiazek}) yields that $W_{j}^{(i)}=\left[  Z_{i},X_{j}\right]  .$ Thus,
due to the fact that we assumed (\ref{Vais}),%
\begin{equation}
L_{Z_{i}}\pi=%
{\displaystyle\sum\limits_{j=1}^{k}}
\left[  Z_{i},X_{j}\right]  \wedge Z_{j}. \label{aaa}%
\end{equation}

\begin{remark}
\label{trivial}In the case that the functions $\varphi_{i}$ are Casimir
functions of $\pi$ we have $X_{i}=\pi d\varphi_{i}=0$ so that the formula
(\ref{aaa}) yields $L_{Z_{i}}\pi=0$ for all $i,$ i.e. the vector fields
$Z_{i}$ are symmetries of $\pi$. In this case our reduction procedure
(\ref{redukcja}) coincides with the standard projection onto a level set of
Casimir functions (symplectic leaf in case there are no other Casimirs apart
from $\varphi_{i}$)\cite{olver}.
\end{remark}

From what we have said above it becomes clear that the above reduction scheme
can be interpreted as a two-step procedure: firstly we deform the original
Poisson tensor $\pi\ $to a Poisson tensor $\pi_{D}$ and then we obtain
$\pi_{R}$ as standard projection of $\pi_{D}$ onto the level set $\mathcal{S}$
of its Casimirs $\varphi_{i}$ (thus we need not calculate the prolongations
$F$ and $G$ in order to define $\left\{  f,g\right\}  _{\pi_{R}}$).

Now we check what can be said about our vector fields $Z_{i}$.

According to Remark \ref{trivial} $L_{Z_{i}}\pi_{D}=0$. On the other hand, due
to (\ref{defstyczna}),
\[
0=L_{Z_{i}}\pi_{D}=%
{\displaystyle\sum\limits_{j=1}^{k}}
\left[  Z_{i},X_{j}\right]  \wedge Z_{j}-\sum_{j=1}^{k}L_{Z_{i}}X_{j}\wedge
Z_{j}-\sum_{j=1}^{k}X_{j}\wedge L_{Z_{i}}Z_{j}%
\]
so that $\sum_{j=1}^{k}X_{j}\wedge\left[  Z_{i},Z_{j}\right]  =0$. Of course
one of the possible realizations of this condition is the case that the
distribution $\mathcal{Z}$ be integrable since then $\left[  Z_{i}%
,Z_{j}\right]  =0$. There are, however, other possibilities here. For example,
if $\left[  Z_{i},Z_{j}\right]  =\sum_{s=1}^{k}c_{ij}^{s}X_{s}$ with
$c_{ij}^{s}=c_{sj}^{i}$, $\sum_{j=1}^{k}X_{j}\wedge\left[  Z_{i},Z_{j}\right]
=0$ as well.

\subsection{The case when $X_{i}$ span $\mathcal{Z}$}

This time we assume that $X_{i}=%
{\textstyle\sum_{k}}
\varphi_{ki}Z_{k}$ for some real valued functions $\varphi_{ij}$, which due to
(\ref{ortogonalnosc}) yields%
\begin{equation}
\varphi_{ij}=%
{\textstyle\sum_{k}}
\varphi_{kj}Z_{k}(\varphi_{i})=X_{j}(\varphi_{i})=\left\{  \varphi_{i}%
,\varphi_{j}\right\}  _{\pi}.\label{fi}%
\end{equation}
The functions $\varphi_{ij}$ define a $k$-dimensional skew-symmetric matrix
$\varphi=\left(  \varphi_{ij}\right)  ,$ $i,j=1,\ldots k$. The only condition
imposed on $\varphi$ is related to the demand that $X_{i}$ span $\mathcal{Z}$,
i.e. $\det\varphi\neq0.$ We thus do not have to assume (\ref{Vais}) this time
since now the distribution $\mathcal{Z}$ is spanned by the Hamiltonian vector
fields $X_{i}$ and thus $\pi$ is automatically invariant with respect to
$\mathcal{Z}$ as $L_{X_{i}}\pi=0$ for all $i$. It can be easily shown that
\[
\lbrack X_{j},X_{i}]=X_{\{\varphi_{i},\varphi_{j}\}_{\pi}}=\pi\,\,d\{\varphi
_{i},\varphi_{j}\}_{\pi}=\pi\,d\varphi_{ij}.
\]

Now we look for solutions of (\ref{warunek}) in the simple form $V_{i}=\alpha
X_{i}$. Inserting this into (\ref{warunek}) and using the fact that
$\varphi_{ij}=-\varphi_{ji}$ we obtain
\[
0=\alpha X_{i}-\alpha\sum_{j=1}^{k}X_{j}(\varphi_{i})Z_{j}-X_{i}=\alpha
X_{i}+\alpha\sum_{j=1}^{k}\varphi_{ji}Z_{j}-X_{i}=(2\alpha-1)X_{i}%
\]
so that $a=1/2$ and $V_{i}=\frac{1}{2}X_{i}$. In this case the deformation
(\ref{deformacja}) attains the form:%
\begin{equation}
\pi_{D}=\pi-\frac{1}{2}\sum_{i=1}^{k}X_{i}\wedge Z_{i} \label{Diracdef}%
\end{equation}
and is, as mentioned above, Poisson. It is easy to check that our operator
$\pi_{D}$ defines the following bracket on $\mathcal{M}$
\begin{equation}
\{F,G\}_{\pi_{D}}=\{F,G\}_{\pi}-\sum_{i,j=1}^{k}\{F,\varphi_{i}\}_{\pi
}(\varphi^{-1})_{ij}\{\varphi_{j},G\}_{\pi}, \label{Diracbracket}%
\end{equation}
where $F,G:\mathcal{M}\rightarrow\mathbb{R}$ are now two \emph{arbitrary}
functions on $\mathcal{M}$, which is just the well known \emph{Dirac
deformation} \cite{Dirac} of the bracket $\{.,.\}_{\pi}$ associated with $\pi$.

\begin{remark}
\label{Diracwlas}If $C:\mathcal{M}\rightarrow\mathbb{R}$ is a Casimir function
of $\pi$, then it is also a Casimir function of $\pi_{D}$ since in this case
(\ref{Diracbracket}) yields
\begin{equation}
\{F,C\}_{\pi_{D}}=\{F,C\}_{\pi}-\sum_{i,j=1}^{m}\{F,\varphi_{i}\}_{\pi
}(\varphi^{-1})_{ij}\{\varphi_{j},C\}_{\pi}=0-0=0. \label{diracbracket}%
\end{equation}
We also know that the constraints $\varphi_{i}$ are Casimirs of the deformed
operator $\pi_{D}$. Thus we can state that Dirac deformation preserves all the
old Casimir functions and introduces new Casimirs $\varphi_{i}$.
\end{remark}

It is now possible to restrict our Poisson operator $\pi_{D}$ (or our Poisson
bracket $\{.,.\}_{\pi_{D}}$) to a Poisson operator $\pi_{R}$ (bracket
$\{.,.\}_{\pi_{R}}$) on the submanifold $\mathcal{S}$, i.e. the level set
$\varphi_{1}=...=\varphi_{m}=0$ of Casimirs of $\pi_{D}$, in a standard way.

\section{Existence of Dirac reduction}

We now present some realizations of the above Dirac case and discuss the
classical concept of the Dirac classification of constraints. We will show
that the classification of constraints as being either of first-class or of
second-class, proposed by Dirac, should be reexamined when one looks at the
problem from a more general point of view.

We recall that a constraint $\varphi_{k}$ is of \emph{first class} if its
Poisson bracket with all the remaining constants $\varphi_{i}$ vanishes on
$\mathcal{S}$, that is if
\begin{equation}
\{\varphi_{k},\varphi_{i}\}_{\pi}|_{\mathcal{S}}=0,\;\;\;\;\;i=1,...,m.
\label{jeden}%
\end{equation}
Otherwise $\varphi_{k}$ is of \emph{second-class}. In the case that at least
one of the constraints is of the first class, the matrix $\varphi_{ij}$ in
(\ref{fi}) is singular on $\mathcal{S}$ so that the formula
(\ref{Diracbracket}) cannot be used in order to define $\pi_{R}$. However, it
may still be possible to define $\pi_{R}$ via the above general scheme. This
indicates that the concept of first class constraint is too narrow. Below we
demonstrate the examples of Dirac reduction in case when constraints
$\emph{are}$ of first class.

We start with a simple example. Consider a $2n$-dimensional manifold
$\mathcal{M}$ parametrized by coordinates $\left(  q_{1,\ldots,}%
q_{n},p_{1,\ldots,}p_{n}\right)  $ and equipped with a Poisson operator of the
form%
\[
\pi=\left[
\begin{array}
[c]{cc}%
0 & Q_{n}\\
-Q_{n} & 0
\end{array}
\right]  ,
\]
where $Q$ is a diagonal matrix of the form $Q_{n}=\mathrm{diag}(q_{1,\ldots
,}q_{n})$. Consider a submanifold $\mathcal{S}$ given by a pair of constraints
$\varphi_{1}(q,p)\equiv q_{n}=0$ and $\varphi_{2}(q,p)\equiv p_{n}=0$. Then
the matrix $\varphi$ has the form%
\[
\varphi=\left[
\begin{array}
[c]{cc}%
0 & q_{n}\\
-q_{n} & 0
\end{array}
\right]
\]
so that it is clearly singular on $\mathcal{S}$ ($\det(S)=0$ on $\mathcal{S}$)
and
\[
\varphi^{-1}=\frac{1}{q_{n}}\left[
\begin{array}
[c]{cc}%
0 & 1\\
-1 & 0
\end{array}
\right]
\]
so that the Dirac formula (\ref{Diracbracket}) cannot be applied. However, the
vector fields $Z_{1}=q_{n}^{-1}X_{2}$ and $Z_{2}=-q_{n}^{-1}X_{1}$ that span
our distribution $\mathcal{Z}$ are not singular on $\mathcal{S}$ since
$X_{1}=-q_{n}\partial/\partial p_{n}$ and $X_{2}=q_{n}\partial/\partial q_{n}$
so that the deformation (\ref{Diracdef}) becomes
\[
\pi_{D}=\pi-\frac{1}{q_{n}}X_{1}\wedge X_{2}=\pi-q_{n}\frac{\partial}{\partial
q_{n}}\wedge\frac{\partial}{\partial p_{n}}=\sum\limits_{i=1}^{n-1}q_{i}%
\frac{\partial}{\partial q_{i}}\wedge\frac{\partial}{\partial p_{i}}%
\]
and is clearly reducible on $\mathcal{S}$. The operator $\pi_{R}$ obtained on
$\mathcal{S}$ parametrized by coordinates $\left(  q_{1,\ldots,}%
q_{n-1},p_{1,\ldots,}p_{n-1}\right)  $ is%
\[
\pi_{R}=\left[
\begin{array}
[c]{cc}%
0 & Q_{n-1}\\
-Q_{n-1} & 0
\end{array}
\right]
\]
This simple example clearly illustrates that Dirac's classification is too
strong. As a second example we consider a particle moving in a Riemannian
manifold $\mathcal{Q}$ of dimension three with a contravariant metric tensor
\[
G=\left[
\begin{array}
[c]{ccc}%
0 & 0 & 1\\
0 & 1 & 0\\
1 & 0 & 0
\end{array}
\right]
\]
given in some coordinates $(q^{1},q^{2},q^{3})$. Suppose that this particle is
subordinated to a holonomic constraint on $\mathcal{Q}$ given by%
\begin{equation}
\varphi_{1}(q)\equiv q^{1}q^{2}+q^{3}=0. \label{wiezy1}%
\end{equation}
This defines a submanifold of $\mathcal{Q}.$ The velocity $v=\sum_{i=1}%
^{3}v^{i}\partial/\partial q^{i}$ of this particle must then remain tangent to
this submanifold so that%

\[
0=\left\langle d\varphi_{k},v\right\rangle =\sum_{i=1}^{3}\frac{\partial
\varphi_{k}}{\partial q^{i}}v^{i}.
\]
and thus in our coordinates $v^{i}=\sum_{j}G^{ij}p_{j}$ the motion of the
particle in the phase space $\mathcal{M}=T^{\ast}Q$ is constrained not only by
(\ref{wiezy1}) but also by the relation%

\begin{equation}
\varphi_{2}(q,p)\equiv\sum_{i,j=1}^{3}G^{ij}\frac{\partial\varphi_{1}%
(q)}{\partial q^{i}}p_{j}\equiv p_{1}+p_{2}q^{1}+p_{3}q^{2}=0 \label{wiezy2}%
\end{equation}
that is nothing else than the lift of (\ref{wiezy1}) to $\mathcal{M}$. The
constraints (\ref{wiezy1})-(\ref{wiezy2}) define a four-dimensional
submanifold $\mathcal{S}$ of $\mathcal{M}$. We now introduce the following
Poisson structure on $\mathcal{M}$:%

\[
\pi=\left[
\begin{array}
[c]{cccccc}%
0 & 0 & 0 & q^{1} & -1 & 0\\
0 & 0 & 0 & q^{2} & 0 & -1\\
0 & 0 & 0 & 2q^{3} & q^{2} & q^{1}\\
-q^{1} & -q^{2} & -2q^{3} & 0 & p_{2} & p_{3}\\
1 & 0 & -q^{2} & -p_{2} & 0 & 0\\
0 & 1 & -q^{1} & -p_{3} & 0 & 0
\end{array}
\right]  .
\]

Again the matrix $\varphi$ is singular, since $\varphi_{12}=2(q^{1}q^{2}%
+q^{3})=2\varphi_{1}$ which obviously vanishes on $\mathcal{S}$. One can,
however, perform the deformation (\ref{Diracdef}). A quite lengthy but
straightforward computation shows that in this case%
\[
\pi_{D}=\left[
\begin{array}
[c]{cccccc}%
0 & 0 & 0 & q^{1} & -1 & 0\\
0 & 0 & 0 & q^{2} & 0 & -1\\
0 & 0 & 0 & -2q^{1}q^{2} & q^{2} & q^{1}\\
-q^{1} & -q^{2} & 2q^{1}q^{2} & 0 & p_{2} & p_{3}\\
1 & 0 & -q^{2} & -p_{2} & 0 & 0\\
0 & 1 & -q^{1} & -p_{3} & 0 & 0
\end{array}
\right]
\]
and this operator can be projected onto $\mathcal{S}$. To do this, one can
first pass to the Casimir variables
\[
(q^{1},q^{2},\varphi_{1}(q),\varphi_{2}(q,p),p_{2},p_{3})
\]
since, due to the fact that it is easiest to eliminate $q^{3}$ and $p_{1}$
from the system of equations $\varphi_{1}=\varphi_{1}(q)=0,$ $\varphi
_{2}=\varphi_{2}(q,p)=0$, we parametrize our submanifold by the coordinates
$(q^{1},q^{2},p_{2},p_{3})$. In these variables the operator $\pi_{R}$ attains
the canonical form
\[
\pi_{R}=\left[
\begin{array}
[c]{cccc}%
0 & 0 & -1 & 0\\
0 & 0 & 0 & -1\\
1 & 0 & 0 & 0\\
0 & 1 & 0 & 0
\end{array}
\right]  .
\]

Our two examples show that the condition (\ref{jeden}) is only a necessary
condition for nonexistence of $\pi_{R}$ on $\mathcal{S}$, but is not a
sufficient one. Hence the definition of first class constraints has to be made
weaker. Even in the case when we deal with a real first class constraint we
can obtain $\pi_{R}$ on $\mathcal{S}$ coming from the Dirac reduction of
$\pi.$ We demonstrate this below.

Firstly we assume that we have a pair of constraints $\varphi_{1},\varphi_{2}$
that define our submanifold $\mathcal{S}=\left\{  \varphi_{1}=0,\varphi
_{2}=0\right\}  $ and such that they are of second class, i.e. that
$\varphi_{12}|_{\mathcal{S}}=\left\{  \varphi_{1},\varphi_{2}\right\}
|_{\mathcal{S}}\neq0.$ It is clear that our submanifold $\mathcal{S}$ can be
parametrized in infinitely many different ways by constraints $\widetilde
{\varphi}_{1}=0$, $\widetilde{\varphi}_{2}=0,$ where
\begin{equation}
\widetilde{\varphi}_{1}=\psi_{1}\varphi_{1}+\psi_{2}\varphi_{2}%
,\ \ \ \ \ \widetilde{\varphi}_{2}=\psi_{3}\varphi_{1}+\psi_{4}\varphi_{2}
\label{nowewiezy}%
\end{equation}
and where $\psi_{i}$ are some functions on $\mathcal{M}$ such that $\psi
_{i}|_{\mathcal{S}}\neq0$ and such that
\begin{equation}
D\equiv\left\vert \frac{D(\widetilde{\varphi}_{1},\widetilde{\varphi}_{2}%
)}{D(\varphi_{1},\varphi_{2})}\right\vert =\psi_{1}\psi_{4}-\psi_{2}\psi
_{3}\neq0. \label{nieznika}%
\end{equation}
One can prove the following

\begin{lemma}
The deformations (\ref{Diracdef}) given by the pair $\varphi_{1},\varphi_{2}$
of constraints and by the pair $\widetilde{\varphi}_{1}$, $\widetilde{\varphi
}_{2}$ of constraints define the same reduced Poisson operator $\pi_{R}$ on
$\mathcal{S}$.
\end{lemma}

\begin{proof}
For the moment we denote the deformation (\ref{Diracdef}) defined through
$\varphi_{1},\varphi_{2}$ by $\pi_{D}$ and the corresponding deformation
defined through $\widetilde{\varphi}_{1}$, $\widetilde{\varphi}_{2}$ by
$\widetilde{\pi}_{D}$. Applying (\ref{Diracdef}) we easily get that for any
two functions $A,B:\mathcal{M}\rightarrow R$
\[
\left\{  A,B\right\}  _{\pi_{D}}=\left\{  A,B\right\}  _{\pi}+\frac{\left\{
A,\varphi_{2}\right\}  _{\pi}\left\{  B,\varphi_{1}\right\}  _{\pi}-\left\{
A,\varphi_{1}\right\}  _{\pi}\left\{  B,\varphi_{2}\right\}  _{\pi}}{\left\{
\varphi_{1},\varphi_{2}\right\}  _{\pi}},
\]
where we have assumed that $\left\{  \varphi_{1},\varphi_{2}\right\}  _{\pi}$
does not vanish on $\mathcal{S}$. Similarly%
\begin{equation}
\left\{  A,B\right\}  _{\widetilde{\pi}_{D}}=\left\{  A,B\right\}  _{\pi
}+\frac{\left\{  A,\widetilde{\varphi}_{2}\right\}  _{\pi}\left\{
B,\widetilde{\varphi}_{1}\right\}  _{\pi}-\left\{  A,\widetilde{\varphi}%
_{1}\right\}  _{\pi}\left\{  B,\widetilde{\varphi}_{2}\right\}  _{\pi}%
}{\left\{  \widetilde{\varphi}_{1},\widetilde{\varphi}_{2}\right\}  _{\pi}},
\label{tilde}%
\end{equation}
where $\left\{  \widetilde{\varphi}_{1},\widetilde{\varphi}_{2}\right\}
_{\pi}$ does not vanish on $\mathcal{S}$ due to (\ref{nieznika}). Using the
relations (\ref{nowewiezy}) between the deformed constraints $\widetilde
{\varphi}_{i}$ and the original constraints $\varphi_{i}$, the Leibniz
property of Poisson brackets and the fact that the functions $\varphi_{i}$
vanish on $\mathcal{S}$ we obtain%
\[
\left.  \left\{  \widetilde{\varphi}_{1},\widetilde{\varphi}_{2}\right\}
_{\pi}\right\vert _{\mathcal{S}}=D\left.  \left\{  \varphi_{1},\varphi
_{2}\right\}  _{\pi}\right\vert _{\mathcal{S}}%
\]
and%
\begin{align*}
&  \left.  \left(  \left\{  A,\widetilde{\varphi}_{2}\right\}  _{\pi}\left\{
B,\widetilde{\varphi}_{1}\right\}  _{\pi}-\left\{  A,\widetilde{\varphi}%
_{1}\right\}  _{\pi}\left\{  B,\widetilde{\varphi}_{2}\right\}  _{\pi}\right)
\right\vert _{\mathcal{S}}\\
&  =D\left.  \left(  \left\{  A,\varphi_{2}\right\}  _{\pi}\left\{
B,\varphi_{1}\right\}  _{\pi}-\left\{  A,\varphi_{1}\right\}  _{\pi}\left\{
B,\varphi_{2}\right\}  _{\pi}\right)  \right\vert _{\mathcal{S}}%
\end{align*}
so that the nonzero terms $D$ in the numerator and denominator of
(\ref{tilde}) cancel and we obtain $\left.  \left\{  A,B\right\}  _{\pi_{D}%
}\right\vert _{\mathcal{S}}=\left.  \left\{  A,B\right\}  _{\widetilde{\pi
}_{D}}\right\vert _{\mathcal{S}}$ which implies that the projections of
$\pi_{D}$ and $\widetilde{\pi}_{D}$ onto $\mathcal{S}$ coincide.
\end{proof}

In this nonsingular case the distribution $\mathcal{Z}$ along which we project
a Poisson tensor $\pi$ usually changes after reparametrization, but
$\mathcal{Z}|_{\mathcal{S}}$ remains the same as can be easily demonstrated.
Thus in case of the second class constraints one has a "canonical" way of
projecting $\pi$ onto $\mathcal{S}$.

We now suppose that the constraints $\varphi_{i}$ are of first class, that is
$\{\varphi_{1},\varphi_{2}\}_{\pi}|_{\mathcal{S}}$ $=0$ and that the
singularity in $\pi_{D}$ is not removable. We may still attempt to define the
projection $\pi_{R}$ by reparametrizing $\mathcal{S}$ as in (\ref{nowewiezy})
above. It turns out that among an infinite set of admissible
reparametrizations there are some exceptional which, although they fulfil the
condition (\ref{jeden}), nevertheless eliminate the singularity in $\pi_{D}$.
In this case, however, by choosing a new parametrization $\widetilde{\varphi
}_{1}$, $\widetilde{\varphi}_{2}$ of $\mathcal{S}$ we change the distribution
$\mathcal{Z}$ even on $\mathcal{S}$ so that we cannot expect that the
projection $\pi_{R}$ will be independent of the choice of the parametrization.
We lose a natural, "canonical" choice of projection, but we still can perform
the projection, although in infinitely many nonequivalent ways. We illustrate
this below in a sequence of examples.

Consider a six-dimensional manifold $\mathcal{M}$ parametrized with
coordinates $(q_{1,}q_{2},q_{3},p_{1},p_{2},p_{3})$ with the following Poisson
operator:%
\[
\pi=\left[
\begin{array}
[c]{cccccc}%
0 & 0 & 0 & 1 & q_{1} & 0\\
0 & 0 & 0 & q_{1} & 2q_{2}+1 & q_{3}\\
0 & 0 & 0 & 0 & q_{3} & 0\\
-1 & -q_{1} & 0 & 0 & -p_{1} & 0\\
-q_{1} & -2q_{2}-1 & -q_{3} & p_{1} & 0 & p_{3}\\
0 & -q_{3} & 0 & 0 & -p_{3} & 0
\end{array}
\right]  .
\]
Consider now a four-dimensional submanifold $\mathcal{S}$ in $\mathcal{M}$
given by the relations%
\begin{equation}
\varphi_{1}(q,p)=q_{3}=0\text{, \ }\varphi_{2}(q,p)=p_{3}=0\text{.}
\label{czyste}%
\end{equation}
It is clear that $\{\varphi_{1},\varphi_{2}\}_{\pi}$ vanishes on the whole
manifold $\mathcal{M}$ (and thus on $\mathcal{S}$) so that these constraints
do not define any Dirac deformation at all. We now deform (\ref{czyste}) as%
\begin{equation}
\widetilde{\varphi}_{1}=\varphi_{1}+\varphi_{2},\ \ \ \ \ \widetilde{\varphi
}_{2}=(-p_{2}-q_{1}p_{1})\varphi_{1}+\varphi_{2} \label{def1}%
\end{equation}
Calculation shows $\left\{  \widetilde{\varphi}_{1},\widetilde{\varphi}%
_{2}\right\}  _{\pi}=(p_{3}-q_{3})q_{3}$ so that $\{\widetilde{\varphi}%
_{1},\widetilde{\varphi}_{2}\}_{\pi}|_{\mathcal{S}}=0$. One can show that
after introducing the Casimir variables $(q_{1},q_{2},\widetilde{\varphi}%
_{1},p_{1},p_{2},\widetilde{\varphi}_{2})$ the deformed operator $\pi_{D}$
attains the form%
\[
\pi_{D}=\left[
\begin{array}
[c]{cccccc}%
0 & 2\frac{q_{1}q_{3}}{q_{3}-p_{3}} & 0 & 1 & -q_{1} & 0\\
-2\frac{q_{1}q_{3}}{q_{3}-p_{3}} & 0 & 0 & q_{1}+2\frac{p_{1}q_{3}}%
{q_{3}-p_{3}} & -q_{1}^{2}+\theta & 0\\
0 & 0 & 0 & 0 & 0 & 0\\
-1 & -q_{1}-2\frac{p_{1}q_{3}}{q_{3}-p_{3}} & 0 & 0 & -p_{1} & 0\\
q_{1} & -q_{1}^{2}-\theta & 0 & p_{1} & 0 & 0\\
0 & 0 & 0 & 0 & 0 & 0
\end{array}
\right]  ,
\]
where now $q_{3}=q_{3}(q,p,\widetilde{\varphi})$ and $p_{3}=p_{3}%
(q,p,\widetilde{\varphi})$ and $\theta=(q_{3}+p_{2}q_{3}+q_{1}q_{3}%
p_{1})/(q_{3}-p_{3})$, and as such is clearly singular on $\mathcal{S}$ and
thus unreducible. This situation seems to be the most common, i.e. a
spontaneous choice of parametrization almost always leads to a singularity.
However, if we perform a slightly different deformation of (\ref{czyste}):%
\begin{equation}
\widetilde{\varphi}_{1}=\varphi_{1},\ \ \ \ \ \widetilde{\varphi}_{2}%
=(-p_{2}-q_{1}p_{1})\varphi_{1}+\varphi_{2} \label{def2}%
\end{equation}
so that $\left\{  \widetilde{\varphi}_{1},\widetilde{\varphi}_{2}\right\}
_{\pi}=-q_{3}^{2}$ is again zero on $\mathcal{S}$, then the operator $\pi_{D}$
becomes nonsingular and its projection on $\mathcal{S}$ has the following form%
\[
\pi_{R}=\left[
\begin{array}
[c]{cccc}%
0 & 0 & 1 & -q_{1}\\
0 & 0 & q_{1} & 1-q_{1}^{2}\\
-1 & -q_{1} & 0 & p_{1}\\
q_{1} & q_{1}^{2}-1 & -p_{1} & 0
\end{array}
\right]
\]
in the variables $(q_{1},q_{2},p_{1},p_{2})$. Yet another deformation (even
this time of the form (\ref{nowewiezy})):
\begin{equation}
\widetilde{\varphi}_{1}=q_{2}\varphi_{1}\text{ \ , }\ \widetilde{\varphi}%
_{2}=(p_{2}+\varphi_{2})\varphi_{1} \label{def3}%
\end{equation}
yields a quite complicated expression on $\left\{  \widetilde{\varphi}%
_{1},\widetilde{\varphi}_{2}\right\}  _{\pi}$:%
\[
\left\{  \widetilde{\varphi}_{1},\widetilde{\varphi}_{2}\right\}  _{\pi
}=(3q_{2}+1)q_{3}^{2}+q_{3}^{3},
\]
so that it again vanishes on $\mathcal{S}$, but $\pi_{D\text{ }}$ is again
nonsingular and in the same variables $(q_{1},q_{2},p_{1},p_{2})$ its
projection becomes%
\[
\pi_{R}=\left[  \begin{displaystyle}
\begin{array}
[c]{cccc}%
0 & 0 & 1-\frac{q_{1}^{2}}{3q_{2}+1} & 0\\
0 & 0 & \frac{q_{1}q_{2}}{3q_{2}+1} & 0\\
\frac{q_{1}^{2}}{3q_{2}+1}-1 & -\frac{q_{1}q_{2}}{3q_{2}+1} & 0 & -\frac
{q_{1}p_{2}}{3q_{2}+1}\\
0 & 0 & \frac{q_{1}p_{2}}{3q_{2}+1} & 0
\end{array} \end{displaystyle}
\right]
\]
which concludes our series of examples.

\section{Conclusions}

In this article we have focused on two issues involving Dirac reductions of
Poisson operators on submanifolds. In the first part of the article we have
shown how the Dirac reduction procedure fits in a natural way, i.e. as a
result of two natural assumptions about the deformation $\pi_{D\text{ }}$ of
$\pi$, in the general Marsden-Ratiu reduction scheme. In the second part of
our considerations we have demonstrated that the Dirac reduction procedure is
often possible even in cases when the constraints that define our submanifold
are of first class (in Dirac terminology), possibly after some suitably chosen
reparametrization of the submanifold $\mathcal{S}$. ~\newline

\noindent\textbf{Acknowledgement} One of the authors (M.B.) would like to
thank Department of Science and Technology, Link\"{o}ping University, for its
kind hospitality during his stay in Sweden in the fall of 2002.

\end{document}